# Martensitic laminate geometry controls electronic phase transitions in a Mott insulator

*Ziming Shao[1], Benjamin Gregory[1], Suchismita Sarker[2], Jacob Ruff[2], Ivan K. Schuller[3], Yoav Kalcheim[4], Andrej Singer[1], ***

[1]Department of Materials Science and Engineering, Cornell University; Ithaca, NY 14853, USA

[2]CHESS, Cornell University; Ithaca, NY 14853, USA

[3]Physics Department, University of California San Diego, La Jolla, California 92093, USA

[4]Department of Materials Science and Engineering, Technion, Israel Institute of Technology; Haifa, 3200003, Israel

*asinger@cornell.edu

Abstract

Symmetry-lowering structural phase transitions result in multiple degenerate structures whose coexistence is determined by macroscopic strain compatibility. In quantum materials, these structural transformations often couple to electronic degrees of freedom, yet how the structural arrangements influence electronic phase transitions remains poorly understood. By analyzing hundreds of diffraction peaks from X-ray reciprocal space mapping, we determine the lattice basis vectors and mutual orientations of all coexisting phases in epitaxial $V_2O_3$ thin films after a symmetry-lowering transformation coincident with a metal-insulator transition. We identify the orientations of interfaces between all coexisting structures using the theory of martensitic phase transformations and find that the low temperature structure comprises finely tuned layered mixtures of alternating twin variants, akin to metal alloys. By comparing films grown on various substrate orientations, we show that the metal-insulator transition temperature increases monotonically with the degree to which these layered mixtures satisfy macroscopic strain compatibility imposed by the substrate.

**Main Text**

*Introduction*

The emergence of electronic or magnetic order often coincides with symmetry-lowering structural transformations. Prominent examples include Mott transitions in complex oxides, such as vanadates[1]. These transformations typically proceed without diffusion through distortions of the unit cell, belong to the class of martensitic transformations, and lead to multiple degenerate variants of the distorted phase. The resulting intrinsic interfaces are a fundamental consequence of the symmetry breaking and emerge via minimization of elastic energy while maintaining crystallographic compatibility between degenerate variants. In epitaxial thin films grown on dissimilar substrates, an additional constraint emerges. During the structural phase transformation, substrate clamping holds the macroscopic in-plane deformation invariant. This constraint alters the microscopic elastic-energy minimization and the crystallographic compatibility between coexisting twins. The substrate orientation determines which crystallographic plane experiences clamping; yet how clamping modulates which twins form and how this structural heterogeneity influences the electronic transition—for example, by shifting the transition temperature—remains unresolved.

The experimental challenge is characterizing all coexisting structures and their interfaces, which requires techniques that capture the full structural complexity. Structural characterization of coexisting phases and their interfaces typically relies on reciprocal space methods or real-space imaging. Electron microscopy offers atomic-scale spatial resolution yet is inefficient in capturing multiple coexisting structures across large volumes. X-ray diffraction can identify twinning via Bragg-peak splitting, but in thin films the diffraction signal is weak due to the small domains (typically tens of nanometers), and a complete crystallographic description requires a measurement of many reflections. Here, we overcome these limitations by applying synchrotron-based large-volume three-dimensional reciprocal space mapping (RSM) to epitaxial $V_2O_3$ thin films to determine all coexisting structures and their interfaces. Using experimental data, we establish martensitic transformation theory as a framework for understanding electronic phase transitions in materials with symmetry-lowering structural distortions.

*Reciprocal space mapping*

Vanadium sesquioxide ($V_2O_3$) is a prototypical Mott insulator[1–4]. It has attracted continued interest for both fundamental studies of strong electron correlations[5–9] and emerging applications in neuromorphic computing[10,11]. We investigate the cooling-induced discontinuous transition from a high-temperature paramagnetic metal (PM) to a low-temperature antiferromagnetic insulator (AFI), which coincides with a symmetry-lowering structural phase transition from a rhombohedral ($R\bar{3}c$, Fig.1A) to a monoclinic ($I2/a$) lattice[12]. This transformation comprises expansion of the basal plane spanned by a=[$2\bar{1}\bar{1}0$]/3 and b=[$\bar{1}2\bar{1}0$]/3, contraction along c, and a tilt of the c-axis around the b axis[1–3,5]. Due to the trigonal symmetry of the parent structure, three structural domains are possible (Fig.1B).

To capture the structural phase transitions within the V$_2$O$_3$ thin films, we collect RSMs (Fig.1C) at both room temperature and low temperature. Figure 1E shows a 2D slice of the 3D RSM (Fig.1D) of the V$_2$O$_3$ thin film grown on C-cut sapphire normal to Q$_z$=[0001] measured at 50 K. The arrangement of Bragg peaks is consistent with the room-temperature data and with the arrangement of the substrate peaks (Fig.S1). Three additional peaks emerge around each reciprocal lattice point of the rhombohedral structure (Fig.1E, inset). These are the Bragg reflections of the three symmetrically equivalent low-temperature monoclinic phases. Our ability to resolve peaks from individual twins while maintaining access to the large reciprocal space enables quantitative crystallographic analysis of all coexisting structures.

*Crystallography of coexisting structures in presence of substrate clamping*

The threefold splitting of peaks in the (0001) plane observed in C-cut films is consistent with bulk crystal behavior[2,12], yet the substrate imposes an additional constraint on thin films. Although the films (~100 nm thick) are relaxed (Fig.S1B), the substrate restricts the macroscopic in-plane extent of the film during the narrow temperature range over which the structural transformation occurs. We will call this constraint substrate clamping. In the C-cut film, this constraint acts in the (0001) plane. The three possible distortions into the lower symmetry structure are degenerate with respect to substrate clamping (Fig.1B). The data are consistent with this symmetry: the Bragg reflections of the three low-temperature twins form an equilateral triangle (Inset in Fig.1B) around the Bragg reflection of the high-temperature rhombohedral phase (Fig.S1) and have approximately equal intensities. Nevertheless, the structural phase transition remains incomplete even at 50 K, far below the structural phase transition temperature of 160 K in bulk, as indicated by the persistent presence of rhombohedral peaks (central peak in inset of Fig.1E). Consistently, the C-cut film remains metallic down to 20 K (Fig.S2). The substrate clamping likely prevents the complete phase transformation. In films grown on A-cut ($11\bar{2}0$), R-cut ($01\bar{1}2$), or M-cut ($10\bar{1}0$) substrates, the monoclinic distortions are no longer equivalent with respect to substrate clamping. Figure 2A-D show the RSMs near the $30\bar{3}6$ Bragg peak of the rhombohedral structure (normal to the ($10\bar{1}2$) planes) for all four substrate orientations, measured at low temperatures. All data show multiple twin peaks, yet their number and arrangement vary significantly with substrate orientation.

Determining lattice constants for coexisting structures is challenging because the peaks from different structures often overlap and the assignment of twin peaks to their corresponding twin structures is unknown. Since our large-volume reciprocal space data provide access to many Bragg reflections, we selectively analyze those where twin peaks are well separated, minimizing ambiguity in peak identification. We then enumerate all possible peak-to-twin assignments combinatorially and determine the optimal solution via linear regression and minimization of root-mean-square error across all peaks simultaneously (Methods). Since twin peaks are well resolved (strain between twins ~0.01) and the detector remains fixed during sample rotation, the relative uncertainty of the real-space basis vectors is $10^{-3}$ or better.

The predicted peak positions from the combinatorial fit show excellent agreement with the experimental measurements (Figs.2A-D). For the C-cut film an additional Bragg peak appears near the position of the room-temperature rhombohedral peak. Based on the fitted structural model, this peak is identified as a low-temperature rhombohedral structure, distinguished by its c-axis remaining normal to the ab-basal plane. Notably, this phase has a longer c-axis and smaller ab-plane area compared to the bulk rhombohedral $V_2O_3$ structure, opposite to pressure-induced isostructural metal-to-insulator transition into the paramagnetic insulating phase. These results are consistent with recent findings based on transport[13], TEM[14,15], and muon-spin rotation[16] of a suppressed and depth-dependent transition into the antiferromagnetic insulating phase in C-cut films.

For the A-cut film, only three coexisting structures are identified at low temperature, corresponding to the three possible monoclinic twins. Although only two peaks are observed near the $30\bar{3}6$ reciprocal lattice point (Fig.2B), the fitted structural model reveals that one of them consists of two overlapping twin Bragg peaks. In the R-cut film, the complex peak configuration observed at low temperature cannot be explained by only three coexisting structures: at least four distinct low-temperature peaks are visible in Figure 2C. By allowing more coexisting structures in the combinatorial fit, we accurately reconstruct the peak configuration with five coexisting structures. All five structures are monoclinic. A comparable level of complexity is observed in the M-cut film at low temperature, where we identified five monoclinic structures. The increased porosity in the R and M-cut samples is a possible reason for additional twin structures (Methods).

After verifying the consistency of the fitted structural model against the experimentally measured Bragg peak configuration and confirming the number of coexisting structures, we now examine the relative structural alignments based on the retrieved structural model. Figure 3 shows the stereographic projections for the four different films centered around [0001]. We present ab-basal plane normals (circles) and c-axis directions (crosses) for each structure in a common triclinic reference frame, which is a distorted rhombohedral lattice (Miller-Bravais indices refer to the parent structure before the monoclinic distortion). For the rhombohedral structure, the stereographic projections of the ab-basal normal and the c-axis determined experimentally overlap, confirming the c-axis is perpendicular to the ab-plane as expected. In contrast, for the monoclinic structures, the stereographic projections of the c-axis and ab-basal normal show a clear deviation, reflecting the rhombohedral to monoclinic distortion (see arrows). The direction of the deviation serves as a distinguishing feature for the three equivalent monoclinic twins distorted along the different in-plane axes.

In the C-cut film (Fig.3A), the ab-basal plane and c-axis of the low-temperature rhombohedral structure align well with those of the high-temperature phase within error margin (Fig.S6). The three monoclinic structures exhibit a three-fold symmetric tilt of their basal planes relative to that of the high-temperature phase, consistent with the RSM shown in Figure 1E, while their c-axes remain closely aligned. This is surprising, as one might expect that the (0001) plane of all twins would align with the (0001) plane of the substrate, which our data shows to be incorrect. In contrast, the A-cut film shows the opposite behavior (Fig.3B): the basal planes of the monoclinic and rhombohedral structures are well aligned, as shown in the A-cut RSM projected along the [0001] direction, whereas the c-axes of the monoclinic structures show a three-fold symmetric tilt relative to the rhombohedral structure. The alignment scheme of the coexisting structures becomes more complex in the R- and M-cut films due to the presence of additional monoclinic structures absent in bulk crystals[12]. Nevertheless, by examining the deviations of the basal plane normals and the c-axes of the five monoclinic structures, we find that there are three unique distortions, similar to the C- and A-cut films. Specifically, structure M4 shares the same distortion with M3 in the R-cut film and M2 in the M-cut film, while M5 exhibits the same distortion as M1 in both cases.

### *Crystallography of interfaces between coexisting structures*

With the experimentally determined lattice bases and their mutual orientations, we can now investigate the crystallography of interfaces between coexisting structures. The theory of martensitic transformations provides two key concepts[17–19]. First, the twin-twin interface (termed the habit plane in martensitic literature[19]): certain crystallographic planes remain undistorted and unrotated by the distortion between two structures, allowing them to match perfectly on that plane and minimize interfacial strain energy. Second, the laminate: individual twins often cannot form low-strain-energy interfaces with the parent phase, and strain minimization can be achieved through finely twinned phase mixtures (Fig.4B). A laminate consists of alternating layers of two twins separated by twin-twin interfaces, characterized by a macroscopic average deformation that depends on their volume fraction. For the laminate to form inside the parent structure, this average deformation must form a low strain-energy interface with the parent phase. We will call the habit plane between twins the "twin-twin interface" and the habit plane between laminate and parent the "laminate-parent interface" (Fig.4B). Building on the experimentally retrieved real-space basis vectors for each structure, we apply the theory of martensitic transformations[19] pairwise to all experimentally determined structures for a particular film (Methods). We identify a stable laminate when three conditions are met: (1) a low-strain twin-twin interface exists, (2) the low-strain laminate-parent interface exists, and (3) the laminate contains nonvanishing volume fractions from both twins (Methods).

The result of this analysis on all samples—C-cut, A-cut, R-cut, and M-cut—shows that all twin-twin interfaces belong to either $(0001)$ or the $\{2\bar{1}\bar{1}0\}$ family of planes (Fig.4A). The C-cut data are reminiscent of the bulk data[12]: all twin-twin interfaces belong to the $\{2\bar{1}\bar{1}0\}$ family of planes. This relationship can be also seen on the stereographic projection in Figure 3A: the vector connecting reciprocal lattice points from different twins points along the normal to the habit plane[19]. The X-ray data confirm presence of $\{2\bar{1}\bar{1}0\}$ interfaces: diffraction streaks along the normal to these planes are visible (Fig.1E, inset)[20]. In the sample grown on C-cut, all twins are equivalent (twin volume fractions in all laminates are about 50%, Table S2) and twin-twin interfaces exist between all twins. The closest plane to form a low-strain interface between a monoclinic twin and the parent phase is from the $\{1\bar{1}00\}$ family, with a twice worse compatibility (Table S2).

In the samples grown on A-cut substrates, the twin-twin interfaces are $(0001)$, in contrast to $\{2\bar{1}\bar{1}0\}$ found in the samples grown on C-cut. While three twins are present, there is no interface between M2 and M3 (Fig.4C, Table S2): these structures do not form a low-strain laminate inside the parent rhombohedral phase. Additionally, only M2 has an interface with the parent rhombohedral phase. In samples grown on R-cut substrates, the M1 and M3 form a laminate; with a twin-twin interface along $(0001)$. The M2 twin forms a direct interface with the rhombohedral parent structure, but no interfaces with either M1 or M3. The two additional twins, M4 and M5, form a twin-twin interface along $(2\bar{1}\bar{1}0)$ (see faint streaks normal to $(2\bar{1}\bar{1}0)$ in Fig.S9). These are the twins reported in [9]. In the samples grown on M-cut sapphire, M1 and M2 form a laminate with a twin-twin interface along $(\bar{1}2\bar{1}0)$. M3 forms an interface with the parent rhombohedral structure. The two additional twins, M4 and M5 form a laminate, with a twin-twin interface along $(0001)$. In all cases, diffraction streaks perpendicular to the identified twin-twin interfaces directly confirm the analysis and corroborate the applicability of the theory of martensitic transformations to $V_2O_3$ (Figs.S7-S10). For films on A, R and M-cut substrates the streaks persist at room temperature after the transformation into the rhombohedral structure, suggesting the interfaces do not entirely vanish after the twins have disappeared.

*Relation of intrinsic strain, substrate clamping, and transition temperature*

The films synthesized on different substrates show different transition temperatures: M-cut has the highest, followed by R-cut and A-cut, while C-cut shows no full transition (Fig.S2). The geometric compatibility of the twin-twin interfaces varies significantly across samples, yet compatibility alone is insufficient to explain the transition temperature differences; both R-cut and M-cut show ideal compatibility between twin pairs, but M-cut transitions at a higher temperature. To understand the temperature dependence, we examine how substrate clamping, which holds the macroscopic in-plane deformation invariant, constrains laminate formation. The laminate-parent morphology is strain-free only on the laminate-parent interface and generally generates strain perpendicular to that plane. Assuming films expand freely normal to the film-substrate interface while remaining clamped in-plane of the substrate, it is plausible that the transformation proceeds most easily when the strain-free laminate-parent interface is parallel to the substrate surface.

Figure 4D shows the misalignment angle between the laminate-parent interface, $\boldsymbol{n}_{\mathrm{LPI}}$, and the substrate-film interface, $\boldsymbol{n}_{\mathrm{S}}$. M-cut shows perfect alignment within experimental accuracy (Table S2): the formation of the laminate for both pairs of coexisting twins (M1-M2 and M4-M5) proceeds with no macroscopic deformation along the film-substrate plane and is fully consistent with macroscopic substrate clamping constraint. R-cut shows approximately 20° misalignment, introducing a small in-plane strain component, and slightly violating the macroscopic substrate clamping constraint. A-cut shows approximately 60° misalignment with yet larger in-plane strain. C-cut produces a laminate-parent interface nearly perpendicular to the substrate normal, placing the strain-free plane normal to the film—the largest in-plane strain emerges, likely preventing the full transition. The key observation of our work is that the degree of required undercooling increases with the mechanical frustration induced by the misalignment between the strain-free laminate-parent interface and the macroscopic strain enforced by the substrate clamping along the film-substrate interface (Fig.4D).

## *Discussion*

In conventional heteroepitaxial strain engineering, substrates impose homogeneous biaxial strain through microscopic lattice matching across a coherent substrate-film interface[21]. Our films are fundamentally different: they are relaxed, with no microscopic coherence between film and substrate. Nevertheless, a strict macroscopic constraint remains: the substrate clamping prevents the film from expanding or contracting in the plane of the substrate surface. This constraint preserves far more structural degrees of freedom during the phase transformation than conventional epitaxial strain engineering. The low temperature structure exploits these additional degrees of freedom by organizing into layered mixtures of alternating twin variants. All twin-twin interfaces belong to either $(0001)$ or $\{2\bar{1}\bar{1}0\}$ families of planes, likely because these planes have the highest cation densities (Methods) and align with the edges of the $VO_6$ octahedra. Most individual twins cannot form low-energy interfaces with the parent structure. Therefore, the laminate-parent interface must be planar[22], a prediction that future transmission electron microscopy experiments can test using the exact interface orientations we report. By comparing films grown on different substrate orientations, we demonstrate that substrate orientation controls the metal-insulator transition temperature through a macroscopic confinement mechanism, reframing it as an extrinsic parameter which can be engineered through substrate selection and film relaxation. We anticipate that our analysis will help resolve longstanding questions about why the transition temperature in $V_2O_3$ depends sensitively on substrate annealing[23,24] or post-synthesis treatments [25].

Martensitic transformations require no diffusion and typically proceed with the speed of sound, which can be on the picosecond timescale in nanomaterials. In Mott insulators these transformations can be driven by optical or electrical stimuli, making them attractive for ultrafast switching[9,26] and neuromorphic computing[10]. Our results suggest that transformation pathways in these complex oxides are reminiscent of transformations in metal alloys[17–19,22]: they are constrained by formation of strain-compatible laminates within the parent structure. Previous work has argued that displacive transformations lead to tweed formation and are reminiscent of spin glasses where frustration can lead to pre-transitional phenomena and delay transformation mechanisms[27]. The persistence of diffraction streaks in the rhombohedral phase suggests that twin boundaries survive into the high-temperature phase, possibly as a result of plastic deformations during the transformation[28,29], providing a structural origin for phase-boundary scarring and memory effects in $V_2O_3$[30]. Incorporating nucleation and growth kinetics into this framework would enable quantitative predictions of transformation pathways and switching timescales, important for memory applications. More broadly, the concepts established here apply to a wide class of correlated oxides that undergo symmetry-lowering structural transitions under macroscopic invariant plane strain constraint, offering a deterministic route to engineer phase behavior through crystallographic compatibility in addition to homogeneous strain.

**Acknowledgments:**

Z.S., Y.K., and A.S. acknowledge support from U.S. - Israel Binational Science Foundation (BSF), under Grant No. 2020337 (thin film synthesis, data collection, theory of martensitic phase transformations). Z.S., B.G., and A.S. acknowledge the support from U.S. Department of Energy, Office of Science, Office of Basic Energy Sciences, under Contracts No. DE-SC0019414 (development of reciprocal space mapping, analysis and interpretation of x-ray data). YK and IKS acknowledge support from the Quantum Materials for Energy Efficient Neuromorphic Computing (Q-MEEN-C), an Energy Frontier Research Center funded by the U.S. Department of Energy, Office of Science, Basic Energy Sciences under Award No. DE-SC0019273 (design of experiment, sample preparation, discussion of the results and writing of the manuscript). This work is based on research conducted at the Center for High-Energy X-ray Sciences (CHEXS), which is supported by the National Science Foundation (BIO, ENG and MPS Directorates) under award DMR-2342336.

**Figures**

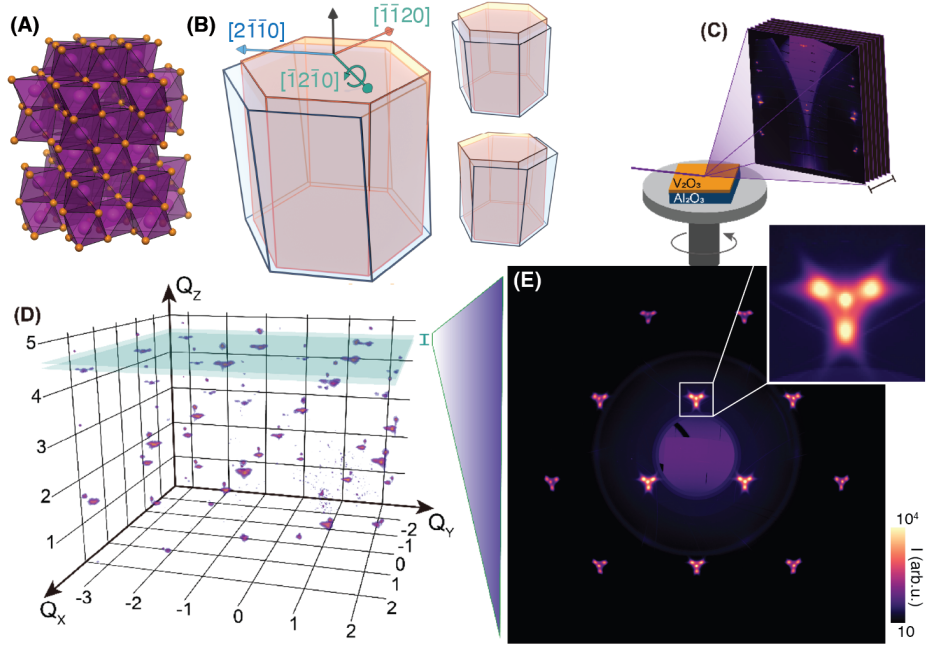

**Figure 1: Reciprocal space mapping. (A)** Conventional hexagonal unit cell of $V_2O_3$ high-temperature rhombohedral phase with oxygen octahedra shown. **(B)** Illustration of lattice distortion during the structural phase transition from the rhombohedral (orange) to monoclinic phase (blue) comprising expansion of basal plane, contraction of c-axis, and rotation of c-axis around b=[$\bar{1}2\bar{1}0$]. The rotation can occur around one of the three in-plane axes: a=[$2\bar{1}\bar{1}0$] (top right), b=[$\bar{1}2\bar{1}0$] (main figure), or -a-b=[$\bar{1}\bar{1}20$] (bottom right). **(C)** Illustration of the large-volume reciprocal space mapping (RSM) experimental set-up. **(D)** Three-dimensional RSM of $V_2O_3$ thin film grown on (0001)-oriented sapphire substrate (C-cut) measured at 50 K. $Q_x$, $Q_y$, and $Q_z$ correspond to the $10\bar{1}0$, $\bar{1}2\bar{1}0$, and 0001 reciprocal lattice directions, respectively. **(E)** Projection of the three-dimensional reciprocal space shown (panel D) over the $Q_z$ range of 4.462 to 4.472 Å$^{-1}$. Inset shows a magnification of the box highlighted in white, in which diffraction from three twins is shown decorated by diffraction streaks due to interfaces.

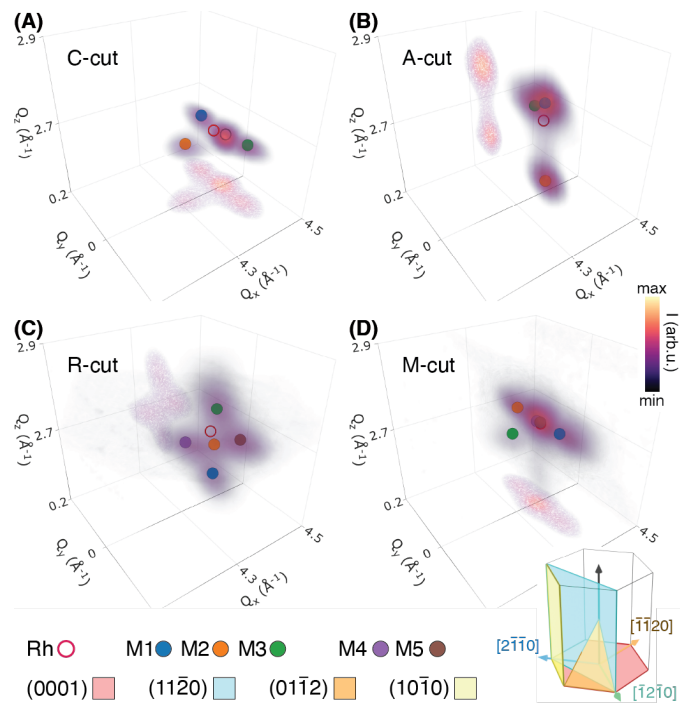

**Figure 2: Peak splitting captured in 3D diffraction intensity. (A-D)** The diffraction intensity near the $V_2O_3$ $30\bar{3}6$ peak for films on C-, A-, R-, and M-cut sapphire substrates shown in logarithmic scale. Selected projections are also shown. Superimposed scatter points denote Bragg peak positions determined from the retrieved structural models: open red circles indicate rhombohedral structures, and closed symbols correspond to monoclinic twins (see legend). Inset in D: The illustration of the common sapphire substrate orientation within hexagonal lattice coordinate.

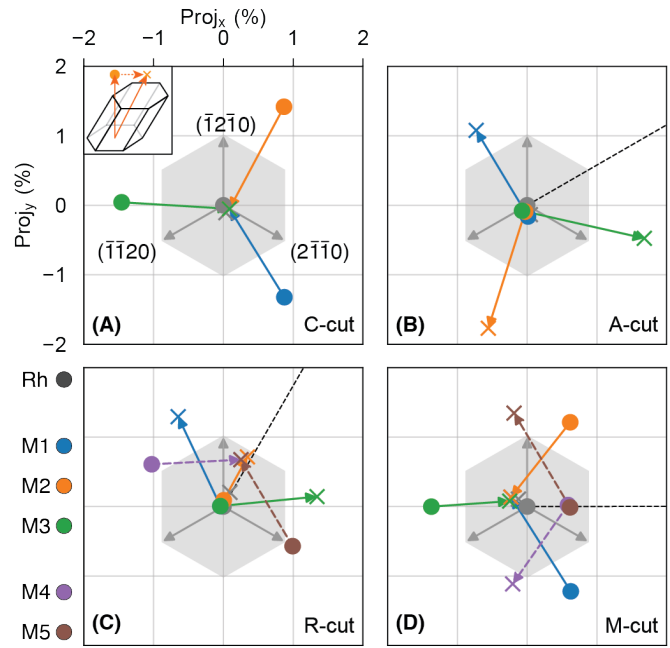

**Figure 3: Experimentally determined stereographic projections. (A-D)** Stereographic projections of the basal plane normal and c-axis directions for coexisting structures in $V_2O_3$ films grown on C-, A-, R-, and M-cut sapphire substrates. Each color denotes a distinct structural type, as labeled in bottom left. Filled circles indicate stereographic projection of the basal plane orientation for each twin. Crosses represent stereographic projections of the c-axis directions. Arrows of matching color connect each pair, illustrating the direction of monoclinic distortion as shown in the inset of panel **A**. The twin numbering is such that all arrows for a certain twin are parallel in **A**-**D**.

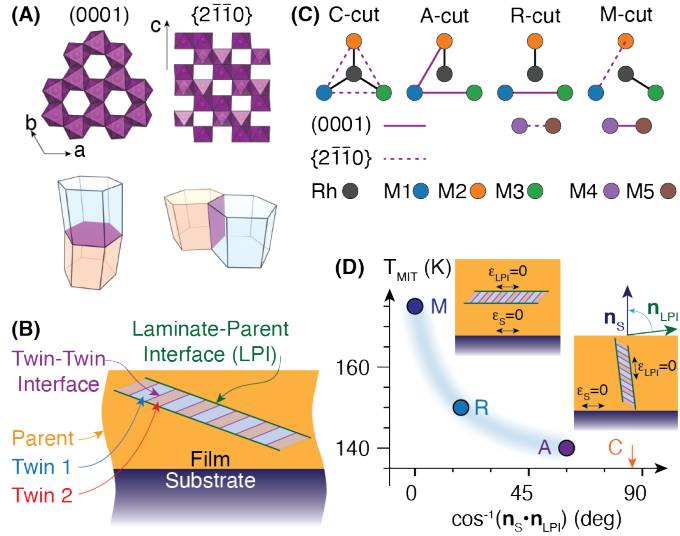

**Figure 4: Results of the theory of martensitic transformations applied to experimental data. (A)** Crystallographic representation of both habit planes observed in the experiment. **(B)** A schematic of the formation of the laminate (mixture two structural low-symmetry variants shown by blue and red), and its invariant plane strain (IPS) with the parent structure (shown in orange). The IPS tilt represents the misalignment of that plane with the substrate plane. **(C)** Schematic representation of all habit planes and the interconnectivity of individual twins measured in each film, grown on C-, A-, R-, and M-cut sapphire. **(D)** The metal insulator transition temperature plotted against the tilt between the laminate-parent interface and the surface normal (see **B**). The insets show the orientation strain-free laminate-parent interface ($\mathbf{n}_{\text{LPI}}$, $\varepsilon_{\text{LPI}} = 0$) with the clamped film-substrate interface ($\mathbf{n}_S$, $\varepsilon_S = 0$). The transition temperature is higher for films with a better alignment between the two interfaces, which reduces structural frustration due to substrate clamping. Films grown on C-cut sapphire have the largest misalignment and show no full transition.

**References**

1. McWhan, D. B., Rice, T. M. & Remeika, J. P. Mott Transition in Cr-Doped V 2 O 3. *Phys. Rev. Lett.* **23**, 1384–1387 (1969).

2. McWhan, D. B. & Remeika, J. P. Metal-Insulator Transition in ( V 1 − x Cr x ) 2 O 3. *Phys. Rev. B* **2**, 3734–3750 (1970).

3. Rice, T. M. & McWhan, D. B. Metal-insulator Transition in Transition Metal Oxides. *IBM J. Res. Dev.* **14**, 251–257 (1970).

4. Imada, M., Fujimori, A. & Tokura, Y. Metal-insulator transitions. *Rev Mod Phys* **70**, (1998).

5. McWhan, D. B., Menth, A., Remeika, J. P., Brinkman, W. F. & Rice, T. M. Metal-Insulator Transitions in Pure and Doped V 2 O 3. *Phys. Rev. B* **7**, 1920–1931 (1973).

6. Limelette, P. *et al.* Universality and Critical Behavior at the Mott Transition. *Science* **302**, 89–92 (2003).

7. Brockman, J. S. *et al.* Subnanosecond incubation times for electric-field-induced metallization of a correlated electron oxide. *Nat. Nanotechnol.* **9**, 453–458 (2014).

8. Alyabyeva, N. *et al.* Metal-insulator transition in V2O3 thin film caused by tip-induced strain. *Appl. Phys. Lett.* **113**, 241603 (2018).

9. Singer, A. *et al.* Nonequilibrium Phase Precursors during a Photoexcited Insulator-to-Metal Transition in V 2 O 3. *Phys. Rev. Lett.* **120**, 207601 (2018).

10. Del Valle, J. *et al.* Subthreshold firing in Mott nanodevices. *Nature* **569**, 388–392 (2019).

11. Del Valle, J. *et al.* Spatiotemporal characterization of the field-induced insulator-to-metal transition. *Science* **373**, 907–911 (2021).

12. Dernier, P. D. & Marezio, M. Crystal Structure of the Low-Temperature Antiferromagnetic Phase of V 2 O 3. *Phys. Rev. B* **2**, 3771–3776 (1970).

13. Kalcheim, Y. *et al.* Structural Manipulation of Phase Transitions by Self-Induced Strain in Geometrically Confined Thin Films. *Adv. Funct. Mater.* **30**, 2005939 (2020).

14. Pofelski, A. *et al.* Domain nucleation across the metal-insulator transition of self-strained $V_2O_3$ films. *Phys. Rev. Mater.* **8**, 035003 (2024).

15. Singh, R. *et al.* Reversal of strain state in a Mott insulator thin film by controlling substrate morphology. Preprint at https://doi.org/10.48550/arXiv.2510.02234 (2025).

16. Huang, C. *et al.* Magnetic Precursor to the Structural Phase Transition in $V_2O_3$. *Adv. Electron. Mater.* **11**, 2500028 (2025).

17. Wechsler, M. S., Lieberman, D. S. & Read, T. A. On the theory of the formation of martensite. *Trans. Met. Soc. AIME* **197**, 1503 (1953).

18. Bowles, J. S. & Mackenzie, J. K. The crystallography of martensite transformations I. *Acta Metall.* **2**, 129–137 (1954).

19. Khachaturyan, A. *Theory of Structural Transformations in Solids*. (Dover Publications, 2008).

20. Warren, B. E. *X-Ray Diffraction*. (Dover Publications, New York, 1990).

21. Haeni, J. H. *et al.* Room-temperature ferroelectricity in strained SrTiO3. *Nature* **430**, 758–761 (2004).

22. Ball, J. M. & James, R. D. Fine phase mixtures as minimizers of energy. *Arch. Ration. Mech. Anal.* **100**, 13–52 (1987).

23. Brockman, J., Samant, M. G., Roche, K. P. & Parkin, S. S. P. Substrate-induced disorder in $V_2O_3$ thin films grown on annealed *c*-plane sapphire substrates. *Appl. Phys. Lett.* **101**, 051606 (2012).

24. Taha Sultan, M., Ignatova, K., Ingvarsson, S. T. & Arnalds, Unnar. B. Influence of substrate orientation and pre-annealing treatment on the metal–insulator transition in V2O3 films: Insights from Al2O3 *r*, *a*, *c*, and *m*-plane substrates. *J. Appl. Phys.* **137**, 045302 (2025).

25. Trastoy, J., Kalcheim, Y., Del Valle, J., Valmianski, I. & Schuller, I. K. Enhanced metal–insulator transition in V2O3 by thermal quenching after growth. *J. Mater. Sci.* **53**, 9131–9137 (2018).

26. Cavalleri, A. *et al.* Femtosecond Structural Dynamics in VO2 during an Ultrafast Solid-Solid Phase Transition. *Phys. Rev. Lett.* **87**, 237401 (2001).

27. Kartha, S., Castán, T., Krumhansl, J. A. & Sethna, J. P. Spin-glass nature of tweed precursors in martensitic transformations. *Phys. Rev. Lett.* **67**, 3630–3633 (1991).

28. Simon, T., Kröger, A., Somsen, C., Dlouhy, A. & Eggeler, G. On the multiplication of dislocations during martensitic transformations in NiTi shape memory alloys. *Acta Mater.* **58**, 1850–1860 (2010).

29. Hull, D. & Bacon, D. J. *Introduction to Dislocations*. (Elsevier, 2011).

30. Vardi, N. *et al.* Ramp-Reversal Memory and Phase-Boundary Scarring in Transition Metal Oxides. *Adv Mater* **29**, 1605029–1605029 (2017).

31. Barazani, E. *et al.* Positive and Negative Pressure Regimes in Anisotropically Strained V2O3 Films. *Adv. Funct. Mater.* **n/a**, 2211801 (2023).

32. Shao, Z. *et al.* X-ray Nanoimaging of a Heterogeneous Structural Phase Transition in V2O3. *Nano Lett.* **25**, 1466–1472 (2025).

33. Veldhuis, S. A., Brinks, P., Stawski, T. M., Göbel, O. F. & Ten Elshof, J. E. A facile method for the density determination of ceramic thin films using X-ray reflectivity. *J. Sol-Gel Sci. Technol.* https://doi.org/10.1007/s10971-014-3336-2 (2014) doi:10.1007/s10971-014-3336-2.

## Methods

### Sample Preparation

We synthesized thin films of $V_2O_3$ approximately 100 nm thick on sapphire, $Al_2O_3$, substrates with four different orientations, C-cut (0001), A-cut (11$\bar{2}$0), R-cut (01$\bar{1}$2), or M-cut (10$\bar{1}$0), by RF magnetron sputtering as described previously in Trastoy, *et al.*[25] and Barazani, *et al*[31].

### X-ray 3D Reciprocal Space Mapping (RSM)

X-ray Diffraction measurements were performed at the Cornell High Energy Synchrotron Source (CHESS) Beamline ID4B. The incident X-ray energy was 15 keV, selected by a double-bounce diamond monochromator. Diffraction patterns were recorded using a Pilatus6M area detector positioned approximately 0.46 meters from the sample.

To construct each three-dimensional reciprocal space map, five scans were conducted to compensate for the detector gaps and capture diffraction in all directions: These include three continuous Phi scans covering 365 degrees, with 0.1° step size and 0.1 second exposure per step, performed at the following angles:

(1) $\chi$=88° and $\theta$ =3°, (2) $\chi$ =90° and $\theta$ =4°, (3) $\chi$ =92° and $\theta$ =5°

Additionally, two theta scans were performed from 0° to 23° in 1000 steps, with 0.1 s exposure per step at the angle of

(4) $\chi$ =90° and $\phi$ =0°, (5) $\chi$ =79° and $\phi$ =0°

To ensure consistent sample alignment, room-temperature and low-temperature measurements for each film were conducted without repositioning the sample. The resulting RSM spans a volume of approximately (5 Å$^{-1}$)$^3$, containing hundreds of Bragg reflections with sufficient resolution to resolve peak splitting from twinning (Fig.1D).

### Combinatorial Lattice Fitting of Coexisting Structures

To determine the lattice constants of coexisting structures in the low-temperature state, we developed a combinatorial fitting algorithm that requires no prior assumption about the relative orientation of different structures, a parameter that is typically unknown due to anisotropic substrate constraints.

We began by identifying Bragg peaks from reciprocal space maps that exhibited clear satellite peak splitting with minimal overlap. Given the large reciprocal space coverage in our measurements, we accessed dozens of Bragg peaks per sample, which enabled us to selectively analyze the most resolvable peaks for structural fitting.

For each selected Bragg peak, the satellite peak positions were extracted by locating local intensity maxima. Due to the peak broadening and partial overlapping, especially at low temperatures, some peak positions may be uncertain or biased. To overcome the uncertainty in peak localization, we developed a combinatorial approach with linear regression: all possible permutations of peak-to-structure assignments were enumerated (see Methods). For each permutation, linear regression was used to solve for the reciprocal lattice basis vectors using an overdetermined set of Bragg peaks per structure. Linear regression is crucial here not only because analytical solutions are unavailable with more than three non-collinear peaks, but also because it provides robustness in determining the reciprocal lattice vectors even when peak positions are imprecise due to the complex Bragg peak configuration.

Then each assignment was evaluated by computing the root mean square error (RMSE) between the experimental peak positions and those reconstructed from the fitted lattice vectors. The final solution was selected based on the assignment that minimized this total RMSE across all structures. This metric effectively penalizes incorrect assignments, as Bragg peaks that do not belong to the same lattice contribute significantly to the total RMSE.

**Stereographic Projection**

To analyze and visualize the relative orientations of the basal plane and c-axis among coexisting structures across different substrate cuts, we employed stereographic projection. This approach maps three-dimensional vectors onto a two-dimensional plane, preserving their angular relationships.

To enable consistent comparison across films with different orientations, each set of vectors of the same cut is aligned such that the basal-plane normal of the room temperature rhombohedral structure point along the +z axis. This alignment preserves the relative angular relationship of the coexisting structures within each film. Then vectors are normalized to unit length, and the stereographic projection is calculated as

$$Proj_X = \frac{x}{1+z}, Proj_Y = \frac{y}{1+z}$$

where (x, y, z) is the normalized vector.

## **Identifying the invariant plane strain**

To identify crystallographically compatible interfaces (habit planes) between coexisting structural domains, we analyze the experimentally determined lattice basis vectors using the geometrical theory of martensitic phase transformations following the procedures in Ref.[19]. For each pair of structures $i$ and $j$ within a given film, we construct the deformation gradient

$$A_{\{ij\}} = \hat{A}_i \hat{A}_j^{-1},$$

where $\hat{A}_i$ and $\hat{A}_j$ are matrices whose columns are the real-space basis vectors of structures $i$ and $j$, respectively. If the transformation between the two structures admits an invariant plane strain (IPS), the deformation can be written as[22,19]

$$A_{ij} = R(I + \boldsymbol{\ell}\boldsymbol{n}^T),$$

where $\boldsymbol{n}$ is the habit-plane normal, $\boldsymbol{\ell}$ is the shear direction, $I$ is the identity, and $R$ is a rigid rotation.

We determined candidate habit-plane normals from the eigenvalue decomposition of the right Cauchy–Green tensor $C = A_{ij} A_{ij}^T$. For each candidate normal $\boldsymbol{n}$, that can be calculated from eigenvectors and eigenvalues of C (see eq. 1.7.13 in Ref. [19]) we quantify the degree of IPS compatibility by projecting the non-identity part of the deformation, $A_{ij} - I$, onto rank-1 dyads of the form $\boldsymbol{a}\boldsymbol{n}^T$. The optimal shear vector is given by $\boldsymbol{a} = (A_{ij} - I)\boldsymbol{n}/(\boldsymbol{n} \cdot \boldsymbol{n})$, and the deviation from ideal rank-1 compatibility is measured by the normalized Frobenius norm

$$\text{res}_{\text{IPS}} = \frac{|(A_{ij} - I) - \boldsymbol{a}\boldsymbol{n}^T|_F}{|A_{ij} - I|_F}.$$

To assess whether pairs of twins can jointly accommodate the parent phase, we extend this analysis to twin laminates. The laminate consists of alternating layers $A_{i0}/A_{j0}/A_{i0}/A_{j0}$ …, where $A_{i0}$ and $A_{j0}$ describe the transformations between the parent phase to the twins $i$ and $j$. The macroscopic deformation of a laminate is characterized by an average

$$\langle A \rangle = x A_{i0} + (1-x) A_{j0},$$

where $x$ is the volume fraction of twin $i$. We determine the optimal laminate composition by minimizing the IPS residual of $\langle A \rangle$ with respect to $x \in [0,1]$. A laminate is considered crystallographically compatible if (i) the twin–twin habit plane admits an IPS with rank-1 residual $\text{res}_{\text{IPS}} < 0.25$, (ii) the optimized volume fraction satisfies $0.1 < x < 0.9$, and (iii) the laminate–parent deformation also admits an IPS with comparably small residual. This procedure identifies twin pairs capable of forming strain-compatible laminates and determines the corresponding habit-plane orientations. We implement this procedure numerically via Python.

**Additional twins in the samples grown on R-cut and M-cut sapphire due to higher porosity.**

The greater freedom in structural alignment observed in films grown on R- and M-cut sapphire may be attributed to the higher film porosity compared to the C- and A-cut. This interpretation is supported by room temperature X-ray reflectivity (XRR) measurements, which show estimated $V_2O_3$ films densities of 4.69, 4.42, 4.20, 3.79 g/cm$^3$ for A-, C-, M-, R-cut films, respectively. We recently reported that films grown on M-cut substrates display a diffraction peak that exists only at intermediate temperatures[32]. We identified this peak as a rotated version of M3, which has an interface with the rhombohedral structure (Fig.S5, Table S2). The data are consistent with a model where twins neighboring voids can rotate without violating compatibility. The rhombohedral structure stabilizes a rotated orientation of M3, producing a peak at intermediate temperatures (Fig.S5); when the rhombohedral structure vanishes, M3 either rotates to coincide with another twin or vanishes.

**The density of cations**

We calculated the density of cations on different high symmetry planes and found the following values:

$\rho_{(0001)} = 9.5\ nm^{-2}, \rho_{\{2\bar{1}\bar{1}0\}} = 10.0\ nm^{-2}, \rho_{\{01\bar{1}2\}} = 7.4\ nm^{-2}, \rho_{\{10\bar{1}0]} = 5.8\ nm^{-2}.$

SUPPORTING INFORMATION
**Martensitic laminate geometry controls electronic phase transitions in a Mott insulator**

*Ziming Shao[1], Benjamin Gregory[1], Suchismita Sarker[2], Jacob Ruff[2], Ivan K. Schuller[3], Yoav Kalcheim[4], Andrej Singer[1, *]*

[1]Department of Materials Science and Engineering, Cornell University; Ithaca, NY 14853, USA
[2]CHESS, Cornell University; Ithaca, NY 14853, USA
[3]Physics Department, University of California San Diego, La Jolla, California 92093, USA
[4]Department of Materials Science and Engineering, Technion, Israel Institute of Technology; Haifa, 3200003, Israel

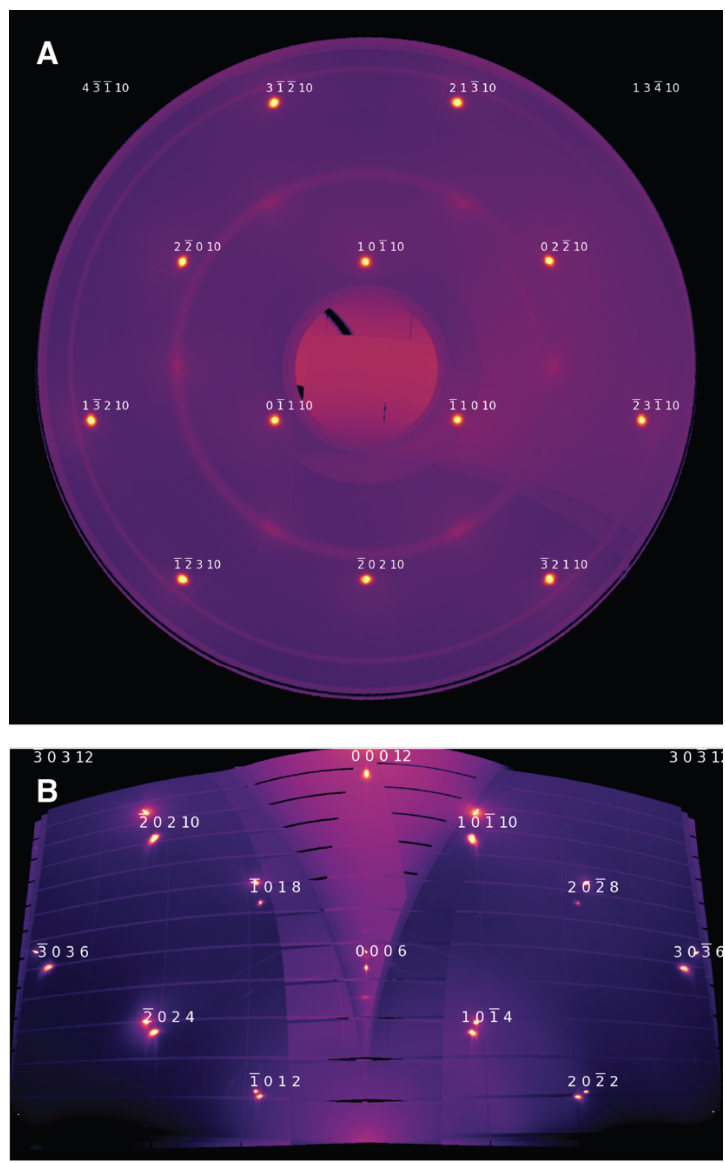

**Figure S1: C-cut RSM 2D Projection at Room Temp.** (A) Projection of the three-dimensional reciprocal space of C-cut film at room temperature over the $Q_z$ range of 4.462 to 4.472 Å$^{-1}$, identical to the range shown in Figure 1E. HKL for each peak is indicated in the four index notation. (B) Projection normal to that shown in A and containing [0001]. The film is relaxed visible by the offset of non-specular film peak from the substrate peak (e.g., the $\bar{3}03\bar{6}$ peak at high temperature).

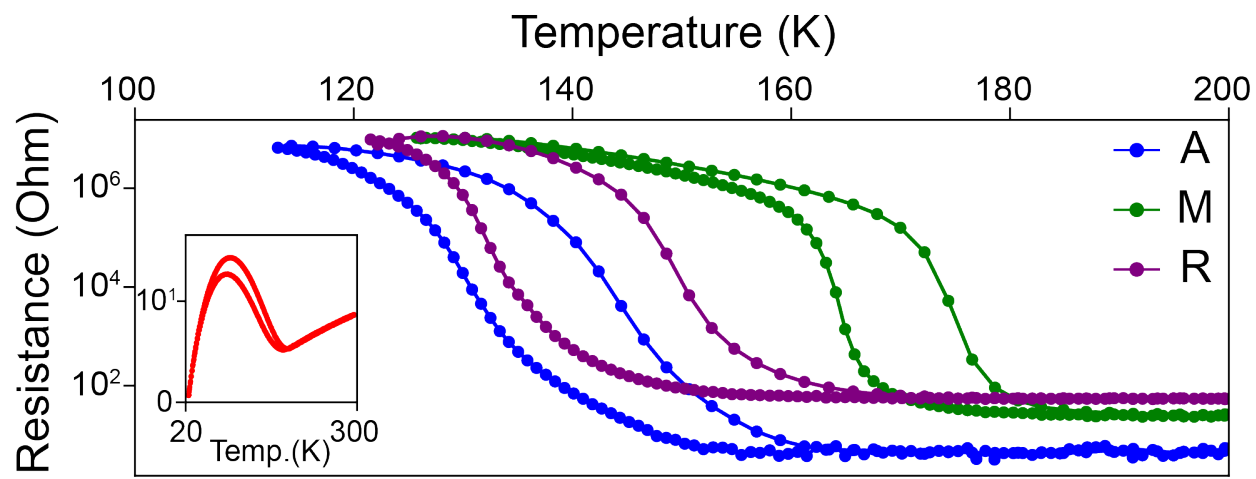

**Figure S2: Temperature Dependent Resistance Measurements.** Resistance as a function of temperature measurements of the $V_2O_3$ thin films grown on sapphire substrates with different orientations. The A-, M-, and R- cuts show obvious metal to insulator transition behavior, while C- cut film (inset figure) shows an increase of resistance at around 150 K, but remains metallic down to 20 K.

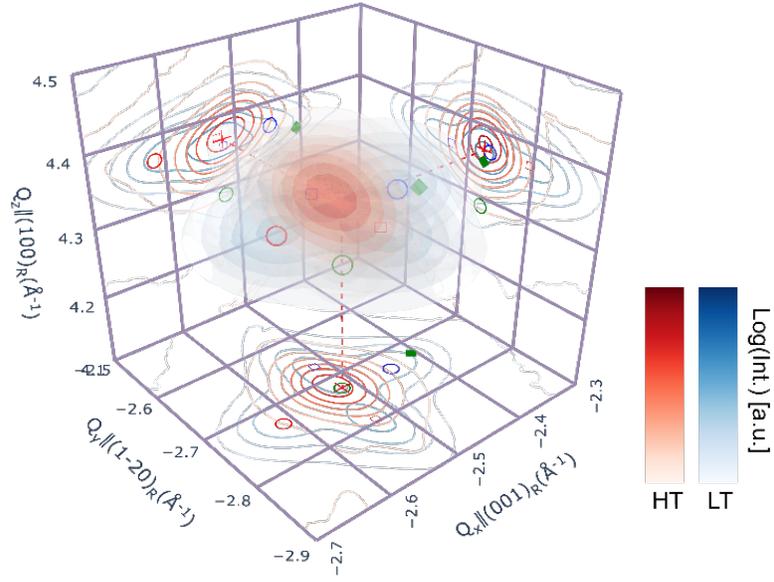

**Figure S3: Complex Low-Temp Peak Configuration for M-Cut Film.** Three-dimensional reciprocal space map near the $V_2O_3$ $22\bar{4}6$ peak for M-cut film, displayed using the same plotting convention as Figure 2 B-E. The complex Bragg peak configuration, indicating more than three coexisting structures, resembles the observation near the $30\bar{3}6$ peak of R-cut film as shown in Figure 2D.

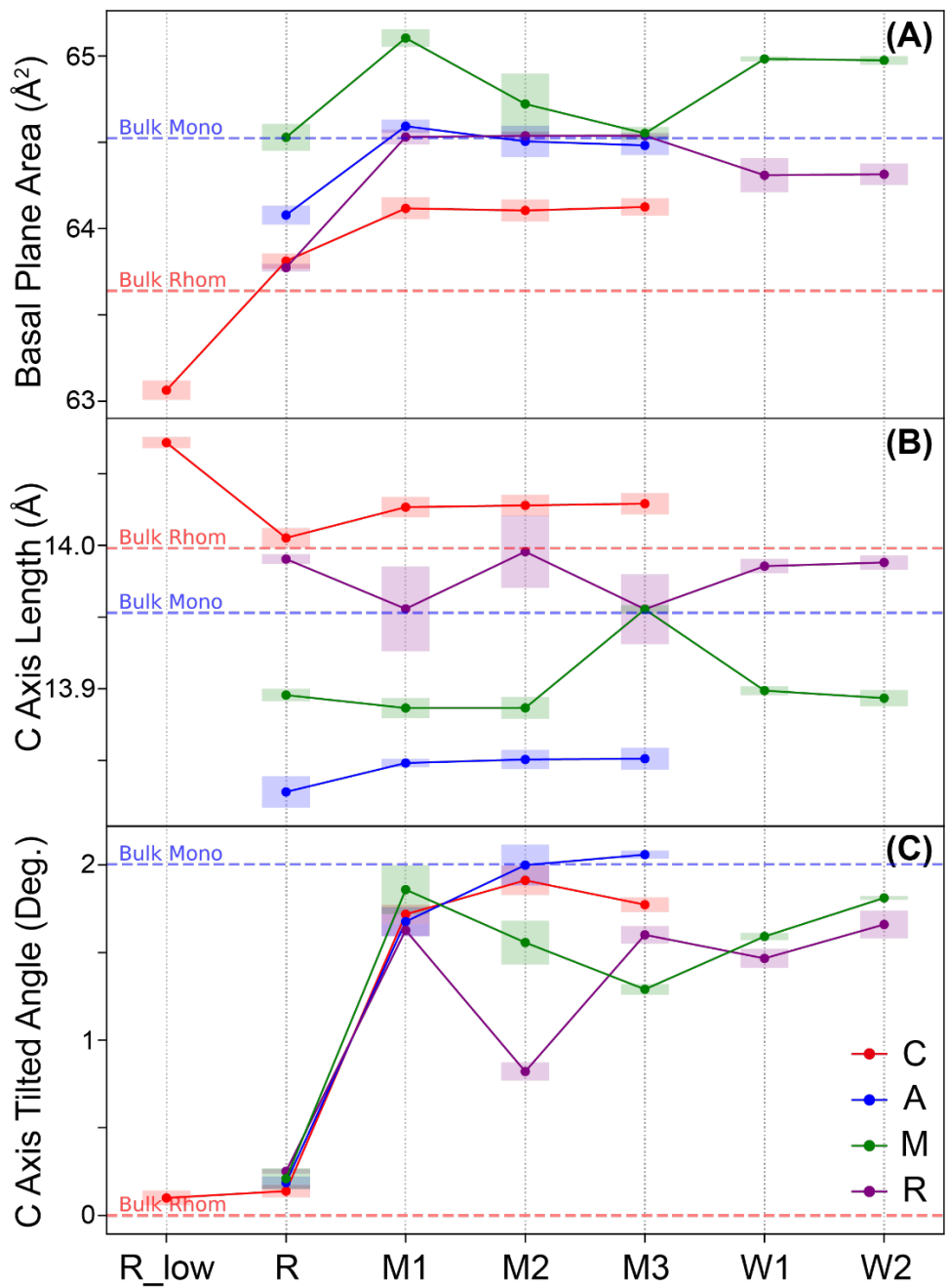

**Figure S4: Comparison of Corundum Unit Cell Shape across Different Structures.** The experimentally determined lattice parameters for all measured films. The parameters expected measured in Bulk are also shown. We parametrize the area of the a-b basal plane (A), c-axis length (B) and the tilt of the c-axis with respect to the a-b basal plane (C). See table S1 for the list of all lattice constants.

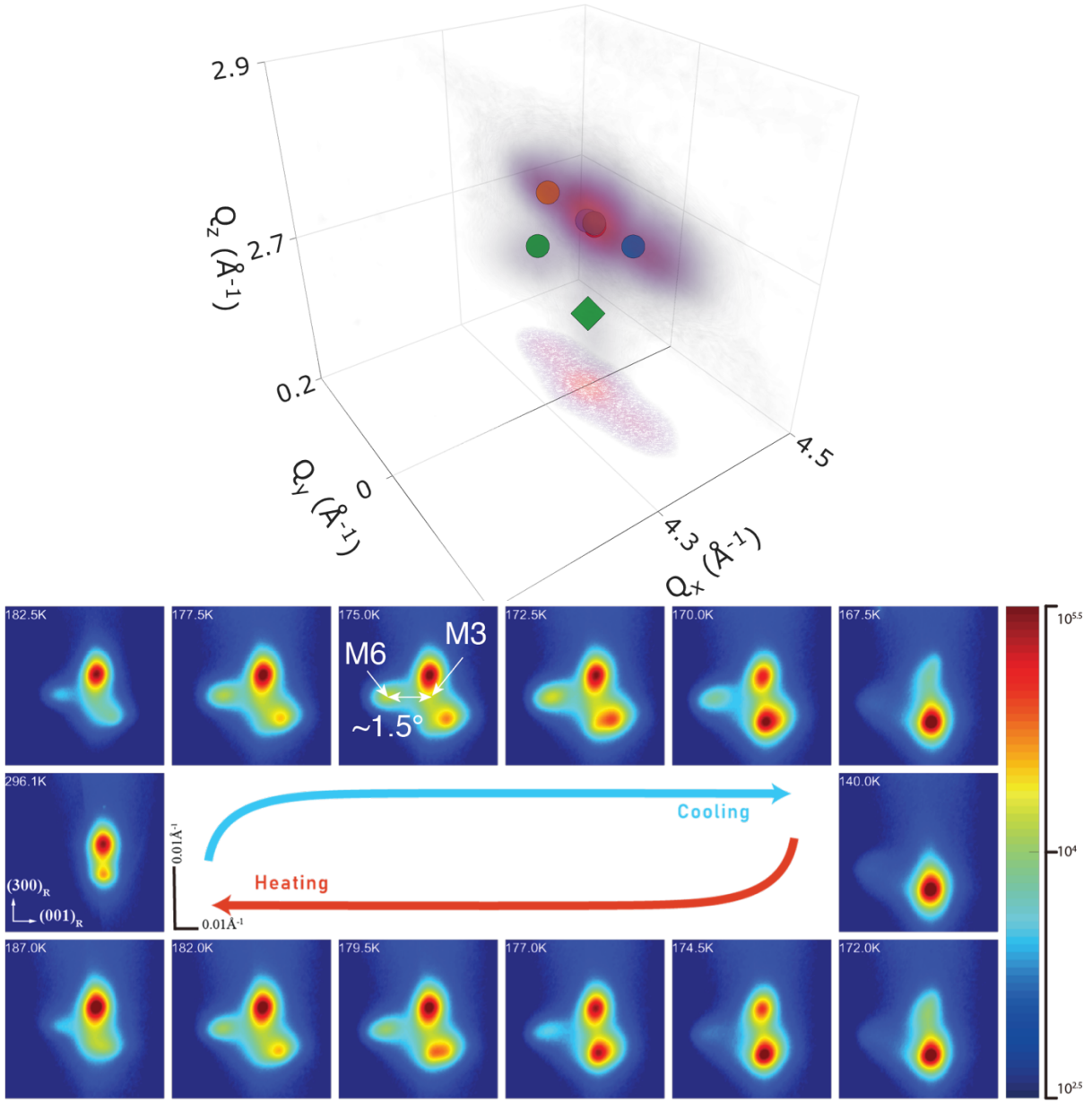

**Figure S5: Explanation for the Intermediate structure reported in Ref.**[32]**.** (Bottom) Figure 2D from the main text with M6 marked as green rhombus. (Bottom) The characterization of the lattice structure observed at intermediate temperatures and reported in Ref.[32], shown here for a 300 nm film grown on M-cut sapphire. The M6 phase is only visible when both, high temperature rhombohedral phase and the low temperature phase are present. We identify the M6 structure as a rotated version of M3 (see Fig.2D of the main text). This twin has an interface with the rhombohedral parent structure (see Fig.4C).

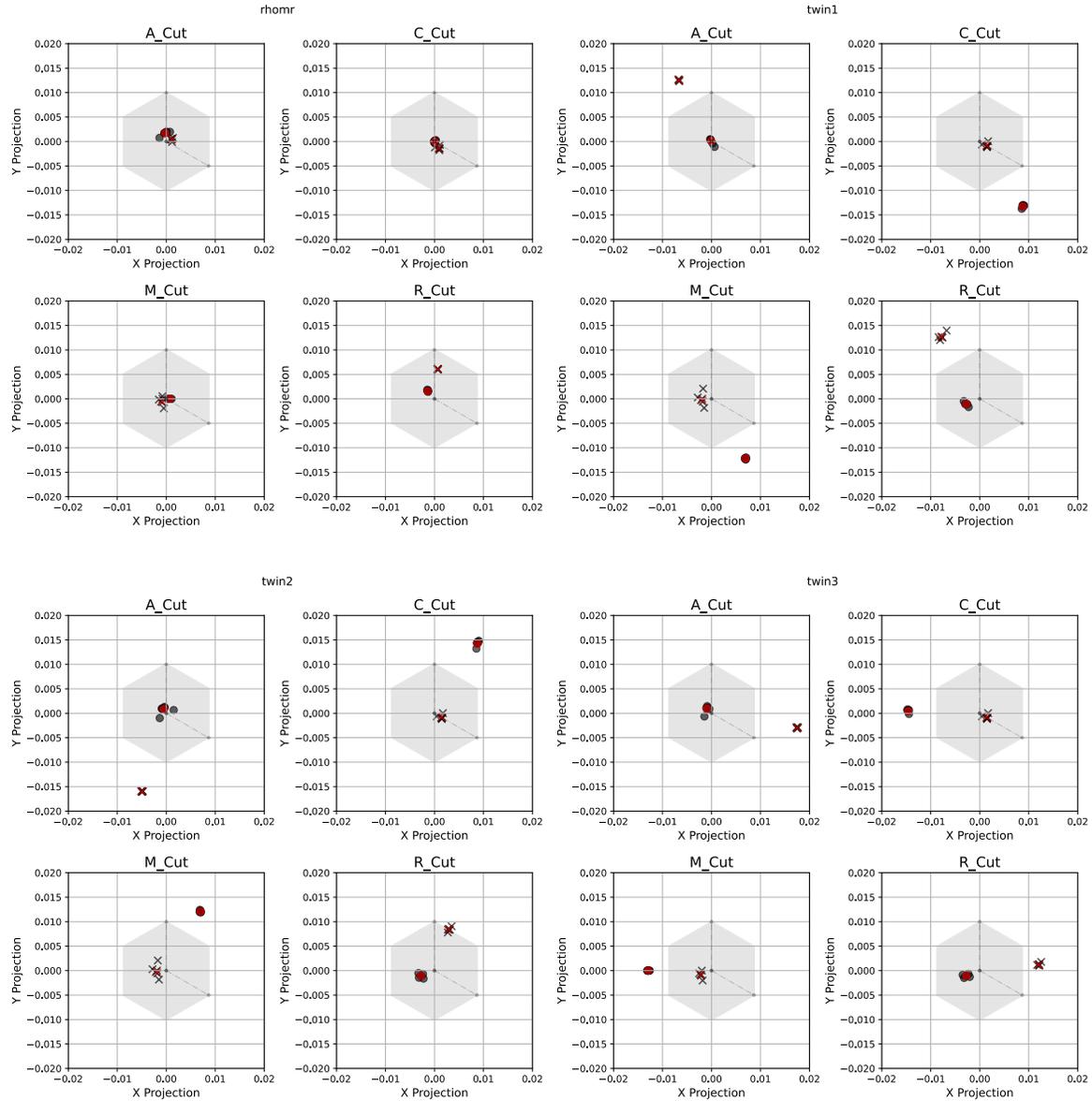

**Figure S6: Error estimation of the stereographic Projection**. The estimate of the uncertainty of the structure determination using a bootstrap method: we calculate the structure parameter by leaving a peak out in the analysis and compare the different results. The spread of the results for each separate analysis can be considered a measure for the uncertainty of the measurements. The figure shows that the spread of the points and crosses for each twin is of the order of the symbol size. All uncertainties are of the same order as the symbols used in the main text. Similar uncertainties are visible in M4 and M5 for samples grown on R and M-cut-substrates.

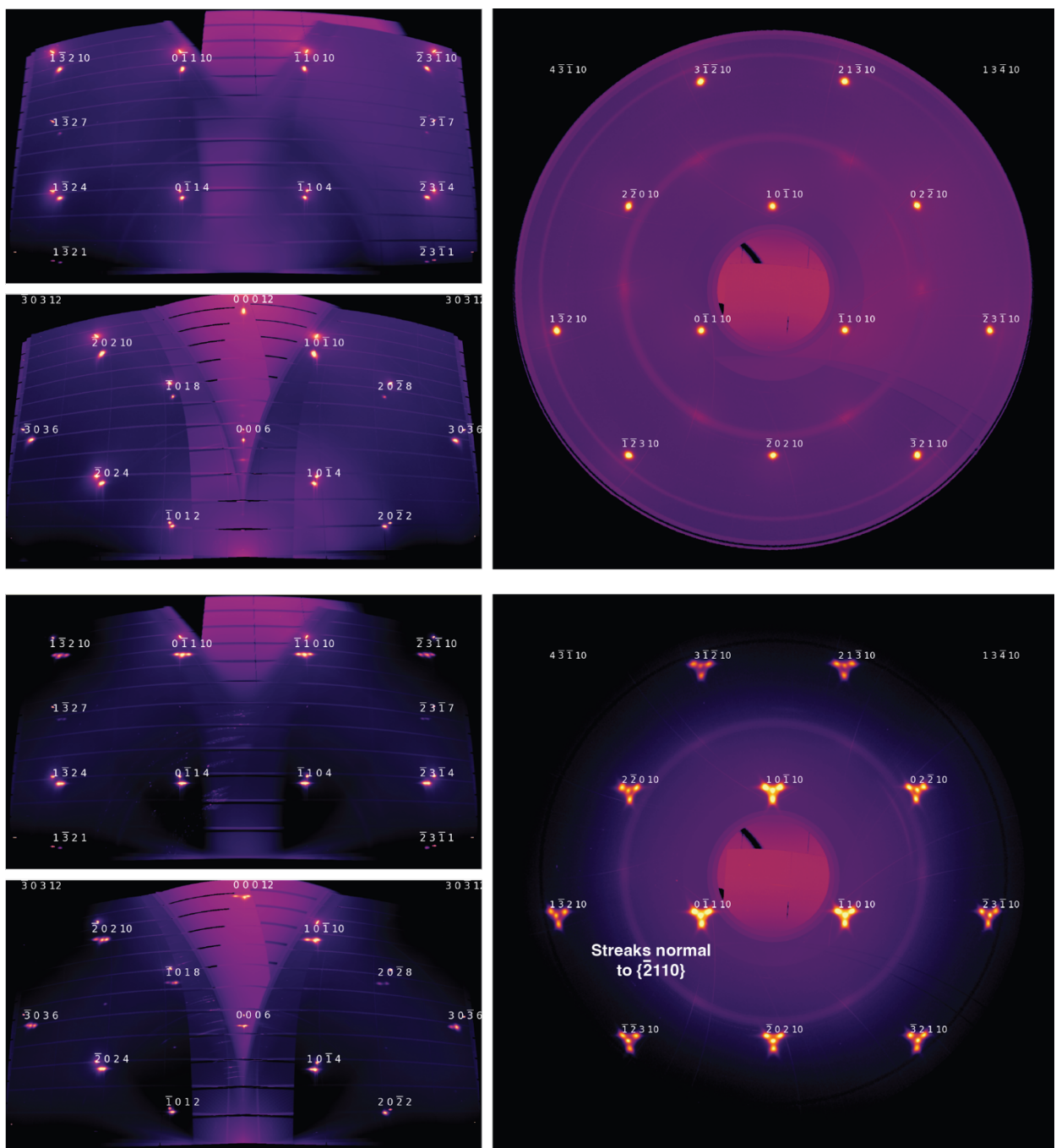

**Figure S7: 2D slices through 3D reciprocal space for V2O3 films grown on C-cut sapphire.** Three orthogonal slices are shown, in laboratory space, i.e., the shown planes are aligned with the substrate normal and edges. The orientation is clear from the HKL notation indicated in each image. The figure shows data collected at 300K (top) and 50 K (bottom). All data was collected using a cryojet. Diffraction streaks, normal to the habit planes are visible at low temperatures. Notice that the films are fully relaxed, i.e., in the film peaks (those with satellites at 50 K) are offset with respect to the substrate peaks along the film normal.

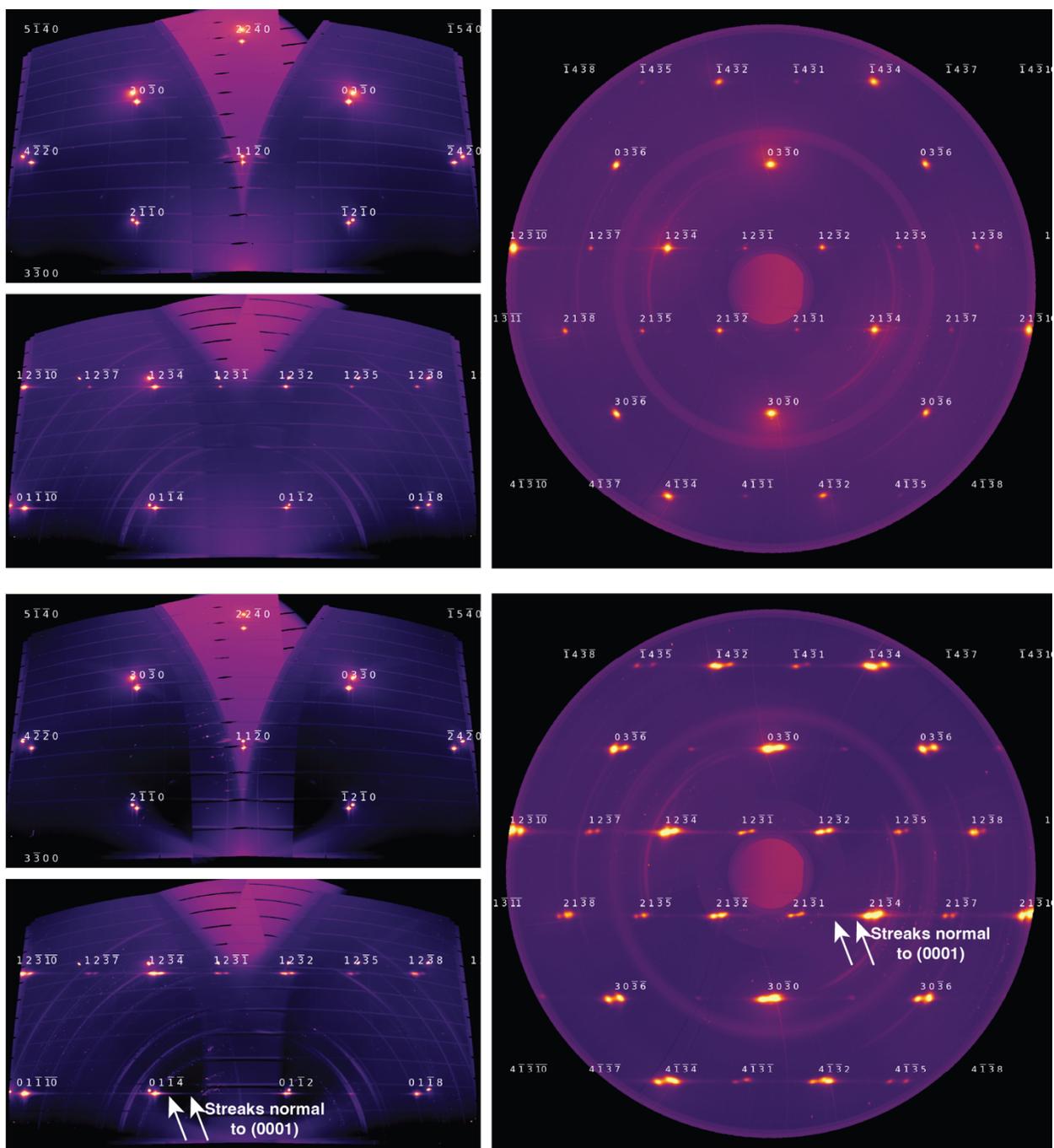

**Figure S8: 2D slices through 3D reciprocal space for V2O3 films grown on A-cut sapphire.** Three orthogonal slices are shown, in laboratory space, i.e., the shown planes are aligned with the substrate normal and edges. The orientation is clear from the HKL notation indicated in each image. The figure shows data collected at 300K (top) and 50 K (bottom). All data was collected using a cryojet. Diffraction streaks, normal to the habit planes are visible at low temperatures. Notice that the films are fully relaxed, i.e., in the film peaks (those with satellites at 50 K) are offset with respect to the substrate peaks along the film normal. Diffraction rings are present due to the scattering on the sample holder and other background. The rings do not affect our analysis.

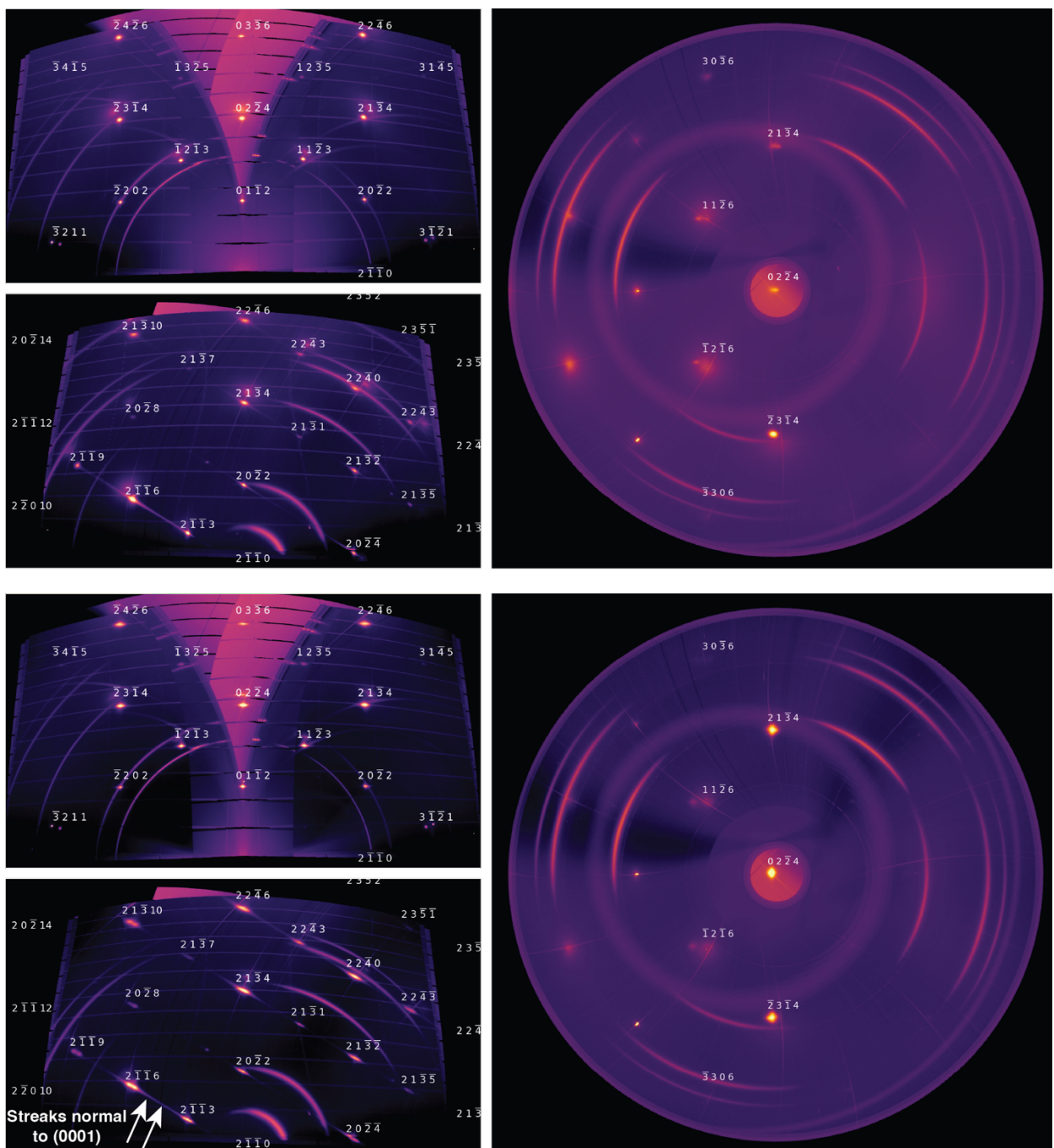

**Figure S9: 2D slices through 3D reciprocal space for V2O3 films grown on R-cut sapphire.** Three orthogonal slices are shown, in laboratory space, i.e., the shown planes are aligned with the substrate normal and edges. The orientation is clear from the HKL notation indicated in each image. The figure shows data collected at 300K (top) and 50 K (bottom). All data was collected using a cryojet. Diffraction streaks, normal to the habit planes are visible at low temperatures. Notice that the films are fully relaxed, i.e., in the film peaks (those with satellites at 50 K) are offset with respect to the substrate peaks along the film normal. Diffraction rings are present due to the scattering on the sample holder and other background. The rings do not affect our analysis.

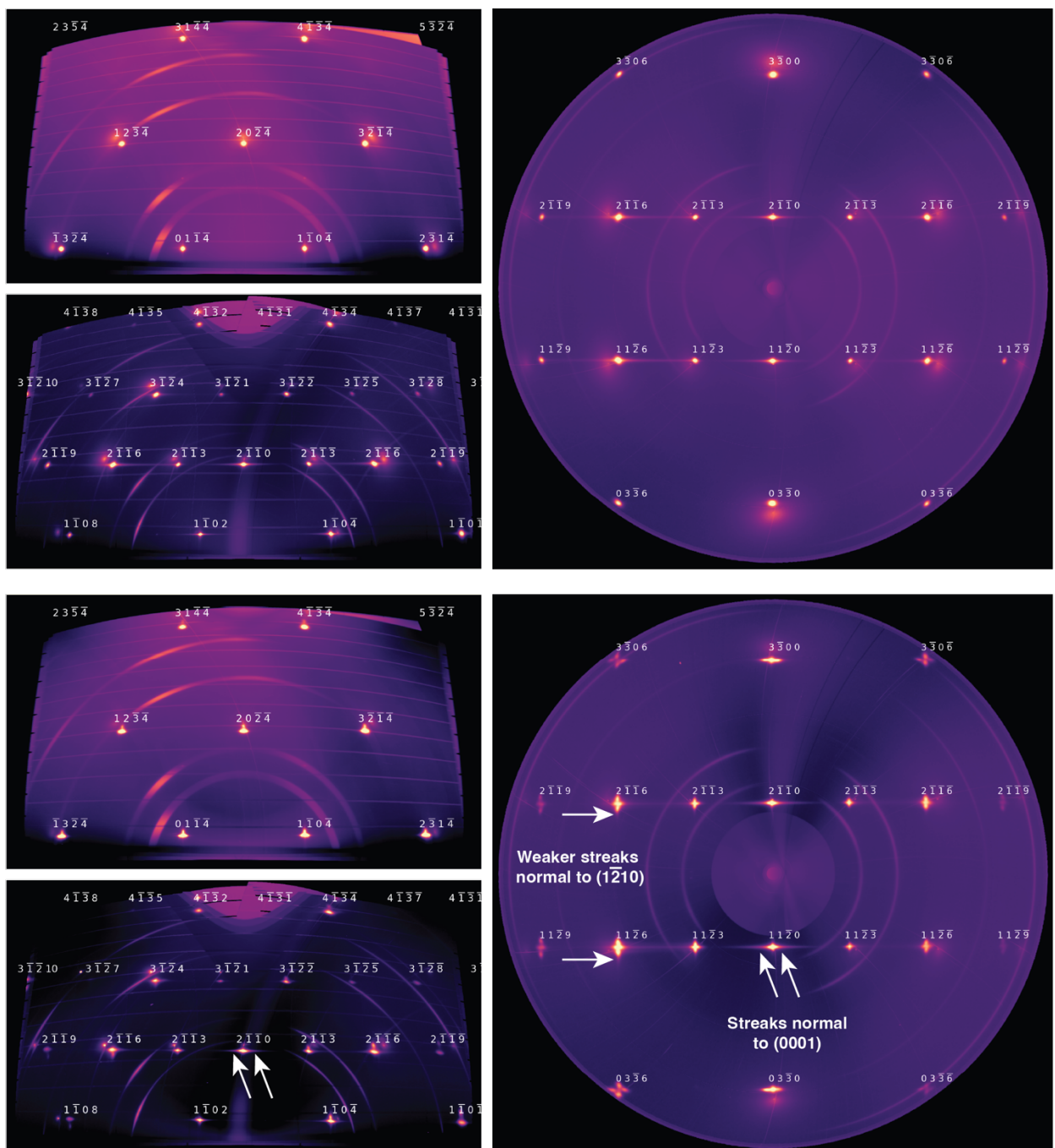

**Figure S10: 2D slices through 3D reciprocal space for V2O3 films grown on M-cut sapphire.** Three orthogonal slices are shown, in laboratory space, i.e., the shown planes are aligned with the substrate normal and edges. The orientation is clear from the HKL notation indicated in each image. The figure shows data collected at 300K (top) and 50 K (bottom). All data was collected using a cryojet. Diffraction streaks, normal to the habit planes are visible at low temperatures. Notice that the films are fully relaxed, i.e., in the film peaks (those with satellites at 50 K) are offset with respect to the substrate peaks along the film normal. Diffraction rings are present due to the scattering on the sample holder and other background. The rings do not affect our analysis.

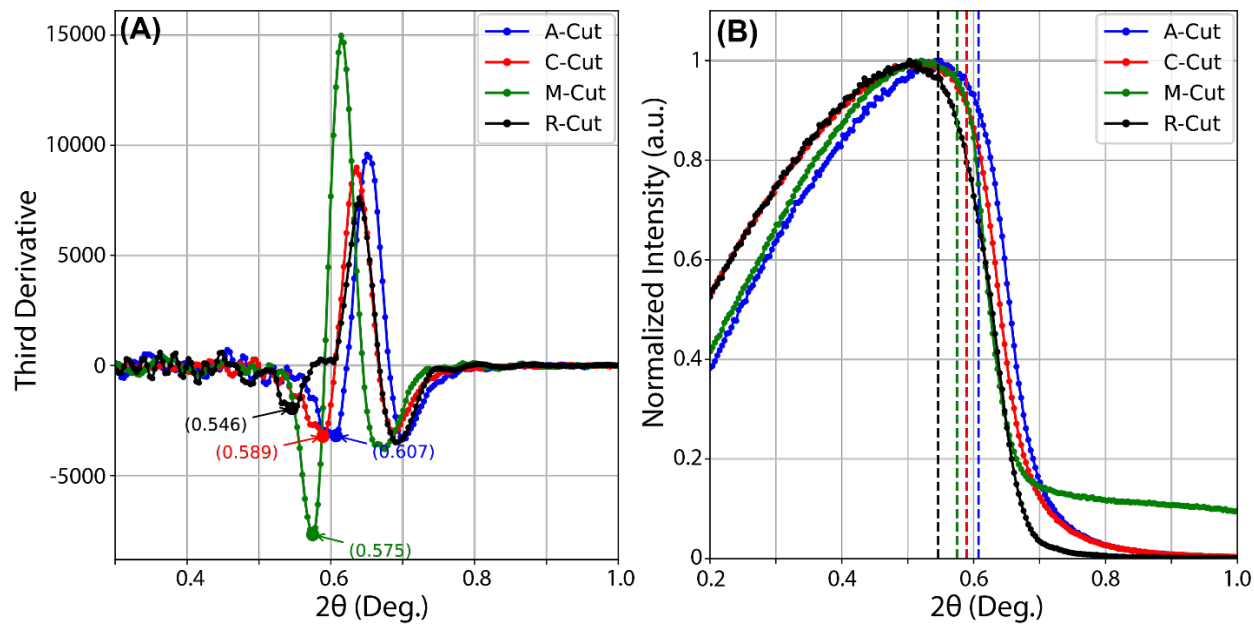

**Figure S11: X-ray Reflectivity (XRR) measurements of $V_2O_3$ films grown on sapphire substrate with different orientations. (A)** The third derivative of the XRR intensity with respect to angle, used to extract pseudo critical angle. This approach is not influenced by the beam footprint, sample size, and surface roughness[13,33]. Moreover, the critical angle can be used to compare the relative density of samples and is shown alongside the normalized XRR curve in panel **(B)**. A clear trend of pseudo critical angle is observed: A-cut > C-cut > M-cut > R-cut, corresponding to the extracted mass density of 4.69, 4.42, 4.20, 3.79 g/cm$^3$, respectively.

| C-cut | a | b | c | γ | β | α |
|---|---|---|---|---|---|---|
| Rh$_l$ | 4.9272 | 4.9310 | 14.0721 | 120.0898 | 89.9827 | 90.0869 |
| Rh | 4.9569 | 4.9578 | 14.0061 | 120.0515 | 89.9040 | 90.1218 |
| M1 | 4.9651 | 4.9634 | 14.0276 | 119.8517 | 91.5077 | 88.5430 |
| M2 | 4.9803 | 4.9620 | 14.0289 | 120.1489 | 89.9344 | 91.6795 |
| M3 | 4.9618 | 4.9780 | 14.0301 | 120.0760 | 88.4133 | 90.1094 |
| **A-cut** | | | | | | |
| Rh | 4.9621 | 4.9541 | 13.8258 | 119.6917 | 89.8782 | 90.1535 |
| M1 | 4.9646 | 4.9634 | 13.8480 | 119.1023 | 91.4400 | 88.5864 |
| M2 | 4.9765 | 4.9525 | 13.8508 | 119.2899 | 89.5827 | 91.9243 |
| M3 | 4.9530 | 4.9775 | 13.8512 | 119.3427 | 88.0327 | 90.4391 |
| **R-cut** | | | | | | |
| Rh | 4.9540 | 4.9519 | 13.9910 | 119.9240 | 90.0236 | 89.7670 |
| M1 | 4.9504 | 4.9954 | 13.9557 | 119.5701 | 91.3369 | 88.5327 |
| M2 | 4.9507 | 4.9992 | 13.9955 | 119.6427 | 90.0203 | 89.2838 |
| M3 | 4.9510 | 5.0027 | 13.9554 | 119.7169 | 88.7044 | 89.8359 |
| M4 | 4.9508 | 4.9838 | 13.9859 | 119.6638 | 88.7686 | 89.9168 |
| M5 | 4.9512 | 4.9829 | 13.9885 | 119.6460 | 91.4456 | 88.5781 |
| **M-cut** | | | | | | |
| Rh | 4.9806 | 4.9836 | 13.8967 | 119.9627 | 90.1854 | 89.8771 |
| M1 | 5.0198 | 4.9815 | 13.8884 | 119.7800 | 91.6232 | 88.4455 |
| M2 | 5.0200 | 4.9514 | 13.8887 | 119.7828 | 90.2370 | 91.2291 |
| M3 | 4.9778 | 4.9836 | 13.9556 | 119.8528 | 88.9288 | 89.9108 |
| M4 | 5.0248 | 4.9716 | 13.8953 | 119.8884 | 90.1413 | 91.2991 |
| M5 | 5.0074 | 4.9728 | 13.8878 | 119.5714 | 91.5667 | 88.4459 |

**Table S1: Crystal structures from large reciprocal space mapping.** We choose a common triclinic reference lattice frame, which is a distorted hexagonal system with $\alpha \approx 90, \beta \approx 90, \gamma \approx 120$ and $a, b, c$ close to the rhombohedral lattice vectors (Table S1). To elucidate the experimentally measured monoclinic distortion we show two projections: ab-basal plane normal (circles) and the c-axis direction (crosses), for each structure, all presented in the common triclinic reference frame. The lattice parameters extracted from large-volume reciprocal space mapping in a common triclinic reference lattice frame. The twins are arranged such that the deformation of $M_j$ across all substrate cuts are similar (for example, the angle beta for all $M_1$ is close to 91.5°). The a, b, c lattice parameters are given in Å, and $\alpha, \beta, \gamma$ are in degrees. The relative uncertainty of all the parameters is of order 10$^{-3}$.

| | Habit plane, structure i vs. j | | | Habit plane, laminate i+j vs. rhomb. | | | | | |
|---|---|---|---|---|---|---|---|---|---|
| **C-cut** | $\mathbf{n}_{TTI}$ | $\|\lambda_2-1\|$ | $\Delta r_{1,TTI}$ | $x$ | $\mathbf{n}_{IPS}$ | $\|\lambda_2-1\|_{IPS}$ | $\Delta r_{1,IPS}$ | $\mathbf{n}_{IPS} \cdot \mathbf{n}_s$ | $\mathbf{n}_{TTI} \cdot \mathbf{n}_{IPS}$ |
| M1–M2 | (-1 2 0) | $5.3\times10^{-5}$ | 0.06 | 0.52 | (-1 -0.1 0.3) | $6.5\times10^{-4}$ | 0.12 | 0.10 | -0.05 |
| M1–M3 | (-2 1 0) | $6.2\times10^{-4}$ | 0.09 | 0.55 | (-0.1 -1 -0.4) | $9.1\times10^{-4}$ | 0.10 | 0.11 | 0.11 |
| M2–M3 | (1 1 0) | $6.7\times10^{-4}$ | 0.08 | 0.48 | (-0.9 1 -0.3) | $1.4\times10^{-3}$ | 0.16 | 0.11 | 0.10 |
| Rh–M1 | (1 -1 -0.2) | $4.4\times10^{-3}$ | 0.15 | | | | | | |
| Rh–M2 | (-0 -1 0.2) | $4.7\times10^{-3}$ | 0.15 | | | | | | |
| Rh–M3 | (1 0 0.2) | $4.1\times10^{-3}$ | 0.15 | | | | | | |
| $R_{LT}$–Rh | (0.4 0.1 -1) | $5.4\times10^{-3}$ | 0.55 | | | | | | |
| $R_{LT}$–M1 | (1 -1 0.3) | $1.1\times10^{-2}$ | 0.34 | 0.29 | (1 -1.0 -0.3) | $1.3\times10^{-3}$ | 0.12 | 0.11 | 0.98 |
| $R_{LT}$–M2 | (0 -1 -0.3) | $1.1\times10^{-2}$ | 0.33 | 0.39 | (-0.2 -2 1) | $5.0\times10^{-4}$ | 0.09 | 0.14 | 0.97 |
| $R_{LT}$–M3 | (1 0 -0.3) | $9.5\times10^{-3}$ | 0.27 | 0.28 | (1 0 0.4) | $1.4\times10^{-3}$ | 0.13 | 0.12 | 0.98 |
| **A-cut** | $\mathbf{n}$ | $\|\lambda_2-1\|$ | $\Delta r-1$ | $x$ | $\mathbf{n}_l$ | $\|\lambda_2-1\|_l$ | $(\Delta r-1)_l$ | $\mathbf{n}_l \cdot \mathbf{n}_s$ | $\mathbf{n} \cdot \mathbf{n}_l$ |
| M1–M2 | (0 0 1) | $1.0\times10^{-4}$ | 0.14 | 0.60 | (-0.1 -0.1 1) | $1.2\times10^{-3}$ | 0.16 | 0.50 | 0.80 |
| M1–M3 | (0 0 -1) | $3.2\times10^{-5}$ | 0.09 | 0.46 | (-0 -0.2 -1) | $1.6\times10^{-3}$ | 0.15 | 0.50 | 0.77 |
| M2–M3 | (0.1 -0.1 -1) | $5.3\times10^{-4}$ | 0.08 | 1.00 | (-0.1 0 1) | $4.1\times10^{-3}$ | 0.16 | 0.22 | -0.98 |
| Rh–M1 | (0 0 1) | $1.0\times10^{-2}$ | 0.36 | | | | | | |
| Rh–M2 | (-0.1 0 1) | $4.1\times10^{-3}$ | 0.16 | | | | | | |
| Rh–M3 | (0 0.1 1) | $5.8\times10^{-3}$ | 0.25 | | | | | | |
| **R-cut** | $\mathbf{n}$ | $\|\lambda_2-1\|$ | $\Delta r-1$ | $x$ | $\mathbf{n}_l$ | $\|\lambda_2-1\|_l$ | $(\Delta r-1)_l$ | $\mathbf{n}_l \cdot \mathbf{n}_s$ | $\mathbf{n} \cdot \mathbf{n}_l$ |
| M1–M2 | (0 0 -1) | $9.2\times10^{-5}$ | 0.07 | 0.00 | (0.1 -0.4 -1) | $6.3\times10^{-4}$ | 0.15 | 0.99 | 0.54 |
| M1–M3 | (0 0 1) | $3.3\times10^{-16}$ | 0.00 | 0.55 | (0.1 -0.3 -1) | $3.4\times10^{-4}$ | 0.17 | 0.94 | -0.70 |
| M2–M3 | (0 0 1) | $9.2\times10^{-5}$ | 0.07 | 0.07 | (0.1 -0.4 -1) | $1.2\times10^{-4}$ | 0.12 | 0.97 | -0.55 |
| M4–M5 | (-2 1 0) | 0 | 0.00 | 0.53 | (-0 -0.2 -1) | $8.6\times10^{-4}$ | 0.23 | 0.96 | 0.07 |
| Rh–M1 | (0.1 0.1 1) | $7.5\times10^{-3}$ | 0.57 | | | | | | |
| Rh–M2 | (0.1 -0.4 -1) | $6.3\times10^{-4}$ | 0.15 | | | | | | |
| Rh–M3 | (-0.1 0.2 1) | $8.2\times10^{-3}$ | 0.48 | | | | | | |
| Rh–M4 | (1 0.1 -0.1) | $6.4\times10^{-3}$ | 0.70 | | | | | | |
| Rh–M5 | (1 -1 0) | $7.5\times10^{-3}$ | 0.66 | | | | | | |
| M1–M4 | (-1 0.6 -0.2) | $3.3\times10^{-3}$ | 0.69 | 1.00 | (0.1 0.1 1) | $7.5\times10^{-3}$ | 0.57 | 0.85 | -0.26 |
| M1–M5 | (-0.1 0.3 1) | $1.9\times10^{-4}$ | 0.86 | 0.80 | (-0.1 -0.1 -1) | $7.6\times10^{-3}$ | 0.55 | 0.84 | -0.89 |
| M2–M4 | (-2 1 0.2) | $3.3\times10^{-3}$ | 0.18 | 1.00 | (0.1 -0.4 -1) | $6.1\times10^{-4}$ | 0.15 | 0.99 | -0.14 |
| M2–M5 | (-2 1 -0.2) | $3.3\times10^{-3}$ | 0.17 | 1.00 | (0.1 -0.4 -1) | $6.3\times10^{-4}$ | 0.15 | 0.99 | -0.10 |
| M3–M4 | (-0.2 0.3 1) | $1.7\times10^{-4}$ | 0.83 | 0.86 | (-0.1 0.2 1) | $8.0\times10^{-3}$ | 0.47 | 0.84 | 0.96 |
| M3–M5 | (0 0 1) | $3.4\times10^{-3}$ | 0.70 | 1.00 | (-0.1 0.2 1) | $8.2\times10^{-3}$ | 0.48 | 0.85 | 0.80 |
| **M-cut** | $\mathbf{n}$ | $\|\lambda_2-1\|$ | $\Delta r-1$ | $x$ | $\mathbf{n}_l$ | $\|\lambda_2-1\|_l$ | $(\Delta r-1)_l$ | $\mathbf{n}_l \cdot \mathbf{n}_s$ | $\mathbf{n} \cdot \mathbf{n}_l$ |
| M1–M2 | (-1 2 0) | $2.2\times10^{-16}$ | 0.00 | 0.49 | (-1 0.0 -0.1) | $3.8\times10^{-3}$ | 0.27 | 1.00 | 0.01 |
| M1–M3 | (-1 0.5 -0.3) | $5.5\times10^{-4}$ | 0.15 | 1.00 | (-1 0.0 -0.6) | $8.7\times10^{-5}$ | 0.10 | 0.99 | 0.87 |
| M2–M3 | (-1 -0.9 -0.6) | $1.1\times10^{-5}$ | 0.17 | 0.00 | (-1 0.0 -0.6) | $8.7\times10^{-5}$ | 0.10 | 0.99 | 0.86 |
| M4–M5 | (-0.1 0 -1) | $2.9\times10^{-4}$ | 0.05 | 0.50 | (-1 -0.1 -0.2) | $2.2\times10^{-3}$ | 0.19 | 1.00 | 0.21 |
| Rh–M1 | (-1 0.8 -0.1) | $5.7\times10^{-3}$ | 0.36 | | | | | | |
| Rh–M2 | (-0.4 -1 -0.4) | $5.7\times10^{-3}$ | 0.43 | | | | | | |
| Rh–M3 | (-1 0 -0.6) | $8.7\times10^{-5}$ | 0.10 | | | | | | |
| Rh–M4 | (0.1 -0.1 -1) | $8.2\times10^{-3}$ | 0.66 | | | | | | |
| Rh–M5 | (0.1 0.1 -1) | $6.9\times10^{-3}$ | 0.59 | | | | | | |
| M1–M4 | (0 0 1) | $1.7\times10^{-4}$ | 0.70 | 1.00 | (-1 0.8 -0.1) | $5.7\times10^{-3}$ | 0.36 | 0.63 | -0.08 |
| M1–M5 | (-0.6 1 0.8) | $1.7\times10^{-4}$ | 0.73 | 1.00 | (-1 0.8 -0.1) | $5.7\times10^{-3}$ | 0.36 | 0.63 | 0.79 |
| M2–M4 | (-0.2 0.9 -1) | $7.5\times10^{-4}$ | 0.71 | 0.87 | (-0.4 -1 -0.3) | $6.0\times10^{-3}$ | 0.41 | 0.71 | -0.77 |
| M2–M5 | (1 -2 0.0) | $5.5\times10^{-4}$ | 0.70 | 1.00 | (-0.4 -1 -0.4) | $5.7\times10^{-3}$ | 0.43 | 0.71 | 0.71 |
| M3–M4 | (-1 -0.9 -0.6) | $3.5\times10^{-3}$ | 0.62 | 0.29 | (0.0 0.0 1) | $7.0\times10^{-3}$ | 0.37 | 0.15 | -0.28 |
| M3–M5 | (1 -0.5 0.3) | $1.0\times10^{-3}$ | 0.69 | 0.30 | (-0.0 0.0 -1) | $6.7\times10^{-3}$ | 0.36 | 0.07 | -0.21 |

**Table S2: Results from the theory of martensitic transformations (i.e., invariant plane strain) analysis sorted by the substrate cut.** Twin-twin interface normal $\mathbf{n}_{TTI}$ (Miller indices of the plane shown in hexagonal coordinates, $hkl$ from $hkil$ by omitting $i$), the deviation of the second eigenvector from 1, rank-1 compatibility. For the laminate the table shows volume fraction $x$ of the two twins in the laminate, the normal of the laminate-parent interface $\mathbf{n}_{IPS}$, deviation of the second eigenvector from 1, rank-1 compatibility for the laminate-parent interface, and dot products among different interfaces.